# Illustrating Einstein's special relativity: A relativistic diagram that displays in true values the components of a four vector

Bernhard Rothenstein[1)], Stefan Popescu [2)] and George J. Spix [3)]

1) Politehnica University of Timisoara, Physics Department, Timisoara, Romania
2) Siemens AG, Erlangen, Germany
3) BSEE Illinois Institute of Technology, USA

***Abstract***. *After having shown that the corresponding components of a four vector transform via the same transformation factors as the space-time coordinates of the same event do, we design a relativistic diagram that displays in true values theirs components. One diagram works-for events generated by tardyons whereas a second diagram works for events generated by light signals or photons. We consider both approaching and receding tardyons respectively photons in each case. We also show how the relativistic diagram for radar and photographic detections of moving profiles.*

## 1. Generating four vectors. What do they share in common?

### 1.1 Hunting with tardyons

Consider the following scenario: an observer $\mathbf{O_0}(0,0)$ located at the origin of the rest frame $K(XOY)$ is equipped with a machine gun $G_0(0,0)$ and with a clock $C_0(0,0)$. A target M moves with constant velocity $V$ parallel to the OX axis in its positive direction. The position of M in the K frame is defined at any time *t* by the space coordinates $M(x = r\cos\theta, y = r\sin\theta)$ using both Cartesian (*x,y)* and polar $(r,\theta)$ coordinates .Let K'(X'O'Y') be the rest frame of the target, where its position is defined by the space coordinates $M'(x' = r'\cos\theta', y' = r'\sin\theta')$. The corresponding axes of the two frames are parallel to each other, whilst the OX and O'X' axes are common. A second observer $\mathbf{O}_0^{'}(0,0)$ is at rest in K' and located at its origin O'. He is equipped with an identical machine gun $G_0^{'}(0,0)$ and with a clock $C_0^{'}(0,0)$. When both clocks defined above read *t=t'=0* the axes of the two frames overlap each other. Observer $O_0'$ orientates the axis of his machine gun along a direction $\theta'$ relative to the positive direction of the common axes and triggers his machine gun at *t'=0* in order to hit the stationary target at time $t' = \frac{r'}{u'}$. Doing so he generates the events $E_0^{'}(0,0,0)$ associated with the triggering of the machine gun and $E'(x' = r'\cos\theta', y' = r'\sin\theta', t' = \frac{r'}{u'})$ associated with the fact that the bullet moving with speed *u'* hits the target.

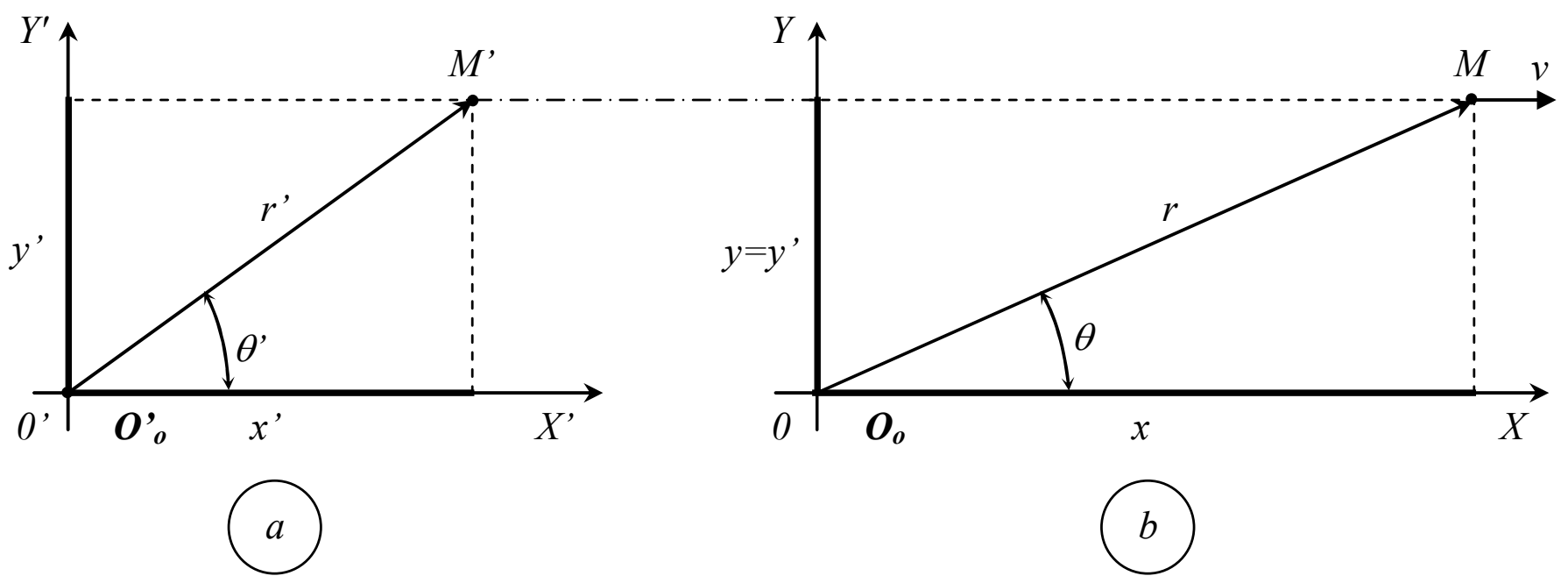

***Figure 1a.*** *Observer* $\mathbf{O}'_0(0,0)$ *located at the origin O' of its rest frame K(XOY) is equipped with a machine gun that fires bullets moving with velocity **u'**. In order to hit a stationary target* $M'(r',\theta')$ *he triggers his machine gun when his clock* $C'_0(0,0)$ *reads t'=0.*

***Figure 1.b.*** *Observer* $\mathbf{O}_0(0,0)$ *located at the origin O of its rest frame K is equipped with an identical machine gun that fires bullets moving with velocity **u.** In order to hit the same target* $M(r,\theta)$ *as* $\mathbf{O}'_0(0,0)$ *above, he triggers his machine gun when his clock* $C(0,0)$ *reads t=0.*

Both $\mathbf{O}_0$ and $\mathbf{O}'_0$ are observers who collect information about events taking place in space either passively (based on the light they receive from point sources i.e. photographic detection) or actively (the observers emit light towards these points in order to locate them i.e. radar type detection)[1,2]. Observer O targets his machine gun along the direction $\theta$ and triggers it at a time *t=0* (event $E_0(0,0,0)$) in order to hit the moving target at position M and time t (event $E(x = r\cos\theta, y = r\sin\theta, \frac{r}{c})$. Having studied special relativity theory, the two observers know that $E$ and $E'$ represent the <u>same event</u>[3] if they take place at the same point in space when the clocks of the two frames K and K' located at the point where the event takes place read *t* and *t*' respectively. This condition is fulfilled if the space-time coordinates of the two events involved in the experiment are related by the Lorentz-Einstein transformations i.e.:

$$x = \gamma r'\left(\cos\theta' + \frac{V}{u'}\right) = D_{x,\theta'} r' \tag{1}$$

$$y = r\sin\theta = y' = r'\sin\theta' \tag{2}$$

$$r = \gamma r'\sqrt{\left(\cos\theta' + \frac{V}{u'}\right)^2 + \gamma^{-2}\sin^2\theta'} = D_{r,\theta'} r' \tag{3}$$

$$r=\gamma^{-1}r'\left[\left(\cos\theta-\frac{V}{u}\right)^2+\gamma^{-2}\sin^2\theta\right]^{-\frac{1}{2}}=D_{r,\theta}r' \quad (4)$$

$$t=\gamma t'\left(1+\frac{V}{c}\frac{u'}{c}\cos\theta'\right)=D_{t,\theta'}t' \quad (5)$$

$$t=\gamma^{-1}t'\left(1-\frac{V}{c}\frac{u}{c}\cos\theta\right)^{-1}=D_{t,\theta}t' \quad (6)$$

$$\tan\theta=\frac{\gamma^{-1}\sin\theta'}{1+\frac{V}{u'}}. \quad (7)$$

Consider a four vector[4,5,] whose vector component is $\mathbf{R}(R_x,R_y)$ and its scalar component is $\Phi$ when detected from K respectively $\mathbf{R}'(R_x',R_y')$ and $\Phi'$ when detected from K'. By definition the corresponding components of the four vector associated with the receding tardyon and with the events it generates transform as the space-time coordinates do i.e.:

$$R_x=D_xR' \quad (8)$$

$$R_y=R_y' \quad (9)$$

$$R=R'D_{r,\theta'} \quad (10)$$

$$R=R'D_{r,\theta} \quad (11)$$

$$\Phi=D_{t,\theta'}\Phi' \quad (12)$$

$$\Phi=D_{t,\theta}\Phi' \quad (13)$$

and of course (7) works in theirs case as well.

At this point we ask our self what significance $R(R')$ and $\Phi(\Phi')$ could have? The answer is that $R(R')$ are the vector components of the *four position vector*, *four velocity, four acceleration, tardyon four momentum*, *four force, four current electromagnetic four potential* as detected from K and K' respectively. Consequently $\Phi(\Phi')$ could be the scalar components of the four vectors mentioned above including the *mass* and *energy* of a tardyon and the *electric charge density*.

### 1.2. Constructing the relativistic diagram that displays in true values the components of a four vector associated with a tardyon as detected from two inertial reference frames in relative motion

The axes of the relativistic diagram we propose are perpendicular to each other and on them we measure (in our two-dimensional approach) $R_x,(R_x')$ and respectively $R_y,(R_y')$. On it we draw the circle of radius $R'$ having its centre at the origin O of the diagram (as shown in Figure 2) where $R'$ represents the magnitude of the vector component of the four vector as measured from K'.

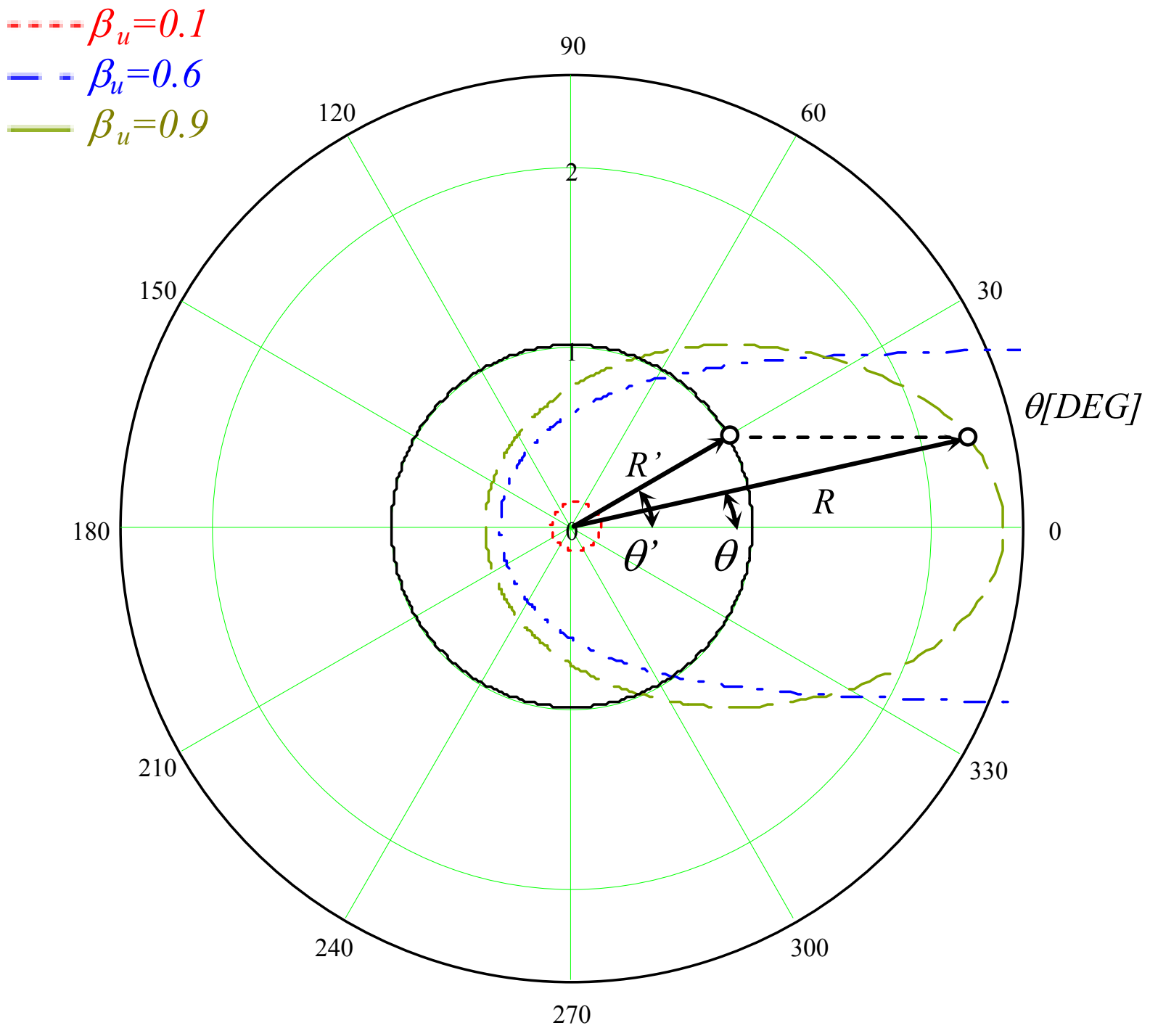

***Figure 2.*** *The relativistic diagram that displays in true values the magnitudes of the space-time coordinates associated with the hunting scenario presented in Figure 1. It also displays in true values the magnitude and the projections of the vector components of four vectors that transform as the space coordinates of the events mentioned above do.*

On the same diagram we draw curve (11). The invariance of the OY(O'Y') components enables us to find out the correspondence between the locations of the end points for the vectors **R** and **R'** as shown in Figure 2. The diagram displays in true values the angles $\theta$ and $\theta'$ which the two vectors make with the positive direction of the common axes. The diagram also depicts at the same scale the following: the circle of radius **R',** the magnitude of **R** and the components of the two vectors obtained by dropping perpendiculars on the corresponding axes.

The second relativistic diagram we propose (presented in Figure 3) displays a circle of radius $\Phi'$ that equals the magnitude of the scalar component of the four vector as measured in K' and the curve (13), $\Phi$ representing its magnitude measured from K. A straight line starting at the origin of the diagram *O* intersects the circle at point **1'** and the curve (13) at point **1.**

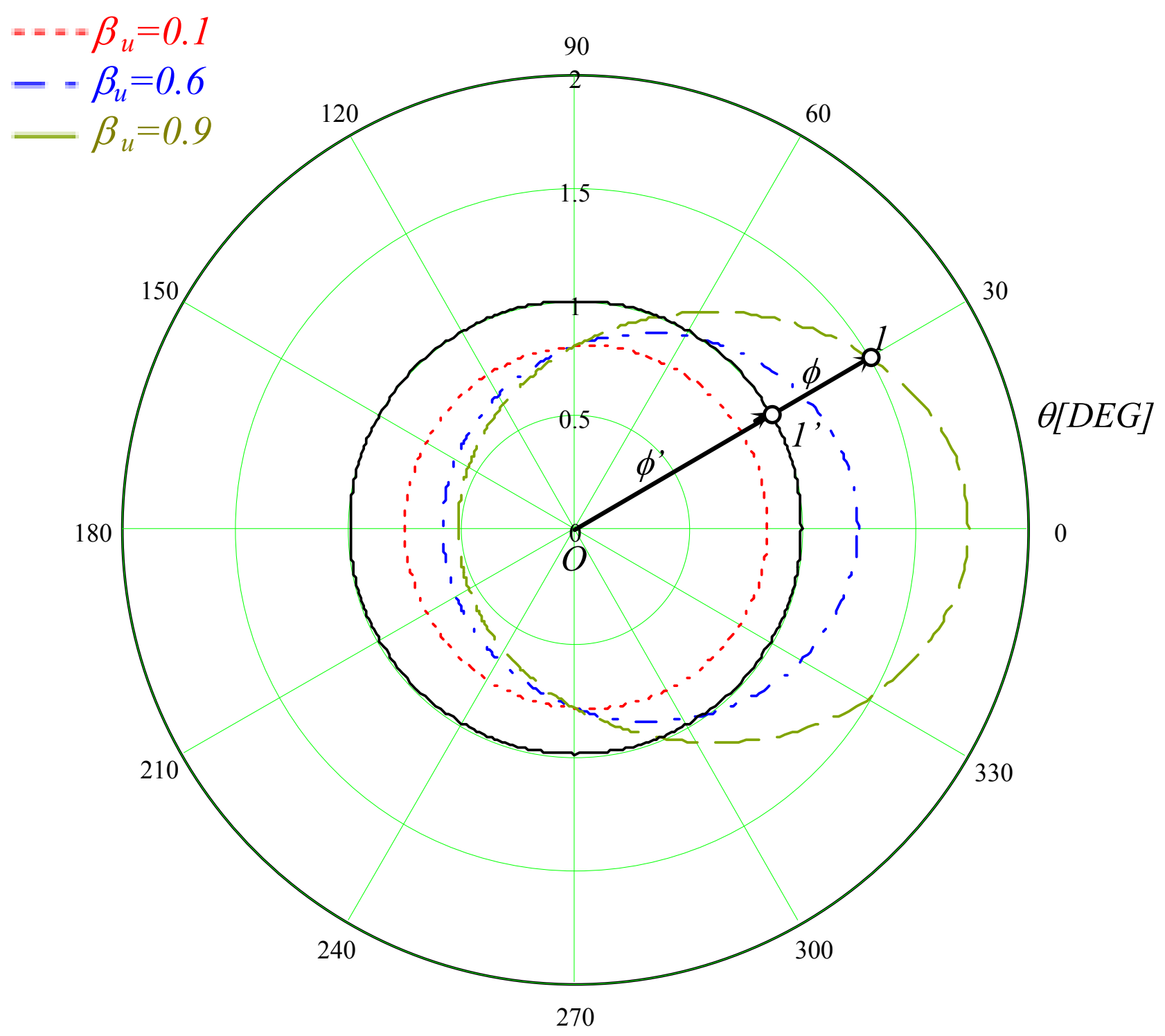

***Figure 3.*** *The relativistic diagram that displays in true magnitudes the scalar components of four vectors that transform like the time coordinates of the events generated by the moving bullets do.*

Based on the geometry of this diagram we find that:

$$\frac{O1}{O1'} = \frac{\Phi}{\Phi'}. \tag{14}$$

Of course after some exercise in handling the diagram we can overlap the two relativistic diagrams presented so far.

In figure 2 as well as in figure 3 we have considered $\beta_V = Vc^{-1} = 0.6$ as well as three different values for $\beta_u = uc^{-1}$ illustrating the way in which the two velocities influence the magnitudes of the physical quantities involved in the transformation process.

### 1.2 Hunting with laser guns[6]

In this case the observers $O_0(0,0)$ and $O_0^{'}(0,0)$ are equipped with laser guns and they hit (illuminate) with light signals the same moving target as in the previous case. The events associated with the hunting in this case are $E'(x' = r'\cos\theta', y' = r'\sin\theta', t' = \frac{r'}{c})$ in K' and $E(x = r\cos\theta, y = r\sin\theta, t = \frac{r}{c})$ in K.

The equations relating the corresponding space-time coordinates are:

$$x = \gamma r'(\cos\theta' + \frac{V}{c}) \tag{15}$$

$$y = r\sin\theta = y' = r'\sin\theta' \tag{16}$$

$$r = \gamma r'(1 + \frac{V}{c}\cos\theta') = D_{c,\theta'} \tag{17}$$

$$r = \gamma^{-1} r'(1 - \frac{V}{c}\cos\theta)^{-1} = D_{c,\theta} r' \tag{18}$$

$$t = \gamma t'(1 + \frac{V}{c}\cos\theta') = D_{c,\theta'} t' \tag{19}$$

$$t = \gamma^{-1} t'(1 - \frac{V}{c}\cos\theta)^{-1} = D_{c,\theta} t'. \tag{20}$$

$$\tan\theta = \frac{\gamma^{-1}\sin\theta'}{\cos\theta' + \frac{V}{c}} \tag{21}$$

As we see, in this case we can transform the position vector magnitudes and the time coordinate magnitudes by the same factor.

Consider the four-vector $(\mathbf{R}_c, \Phi_c)$ in K and $(\mathbf{R}'_c, \Phi')$ in K' whose components transform by definition as:

$$R_{c,x} = D_{c,x} R' \tag{22}$$

$$R_{c,y} = R'_{c,y} \tag{23}$$

$$R_c = D_{c,\theta'} R'_c \tag{24}$$

$$R_c = D_{c,\theta} R'_c \tag{25}$$

$$\Phi_c = D_{c,\theta'} \Phi'_c \tag{26}$$

$$\Phi_c = D_{c,\theta} \Phi'_c \tag{27}$$

The equation (21) is applicable as well.

The vector component of the four vector could be the momentum of a photon $(\mathbf{p_c}, \mathbf{p}'_c)$ and the wave vector of a plane electromagnetic wave $(\mathbf{k}, \mathbf{k}')$ whereas the scalar components could be the energy of a photon $(\varepsilon, \varepsilon')$ and the frequency of the electromagnetic oscillations taking place in the plane wave. In Figure 4 we present the relativistic diagram that displays in true values the vector and the scalar components of the four vectors defined above.

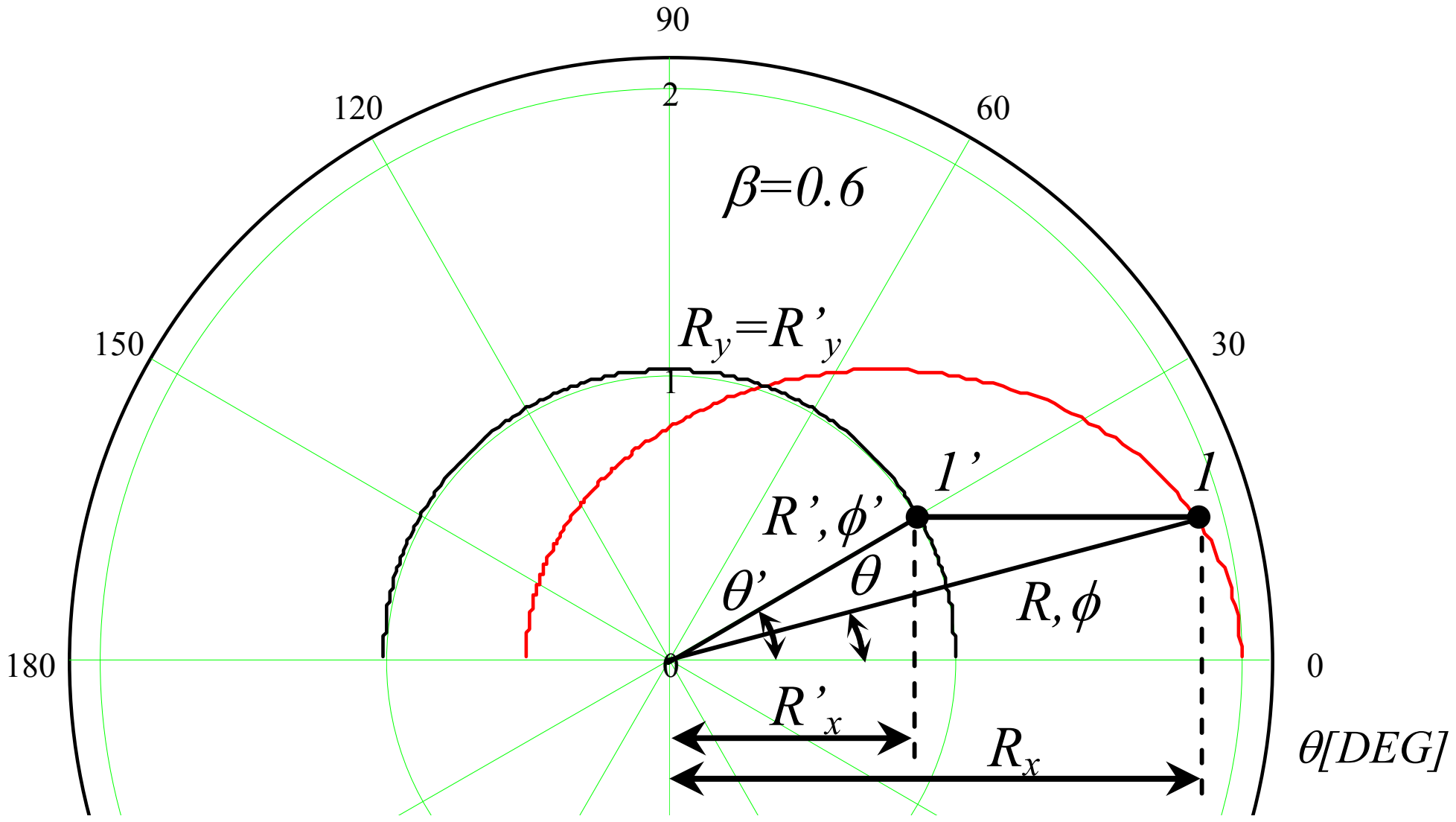

***Figure 4.*** *The relativistic diagram that displays in true magnitudes the space-time coordinates of events associated with the hunting scenario using laser guns. It also displays in true magnitudes the vector and the scalar components of four vectors that transform as the space-time coordinates, mentioned above do.*

As in the previous case we measure on its axes the components of ***R*** and ***R'.*** On it we draw a circle whose radius equals at a given scale the magnitudes of the vector or of the scalar components as measured in K'. We draw also the ellipse described by (25) or (27) at well-known scales. Again, the invariance of the OY(O'Y') and of the vector component enables us to find out in true values on this diagram the relative positions of the corresponding events and the magnitudes of the physical quantities involved.

**2. Simultaneously receiving light signals emitted by distant light sources at different times (photographic detection)**[7]

The scenario we follow involves the observer $O'_0(0,0)$ defined above and a point-like source of light $S'(r',\theta')$ at rest in K'. If the source emits a light signal at $t' = -\frac{r'}{c}$ then the observer $O'_0(0,0)$ will receive it at a zero time. The emission of the light signal is associated with the event $E'(x' = r'\cos\theta', y' = r'\sin\theta', t' = -\frac{r'}{c})$ and its reception at O' is associated with the event $E'_0(0,0,0)$ that has the same space-time coordinates in all inertial reference frames. In K the event *E'* is characterized by the space-time coordinates:

$$x = \gamma r'(\cos\theta' - \beta) = D_{x,-c} r' \quad (28)$$

$$y = r' \sin\theta' \quad (29)$$

$$r = \gamma r'(1 - \beta\cos\theta') = D_{-c,\theta'} r' \quad (30)$$

$$r = \gamma^{-1} r'(1 + \beta\cos\theta)^{-1} = D_{-c,\theta} r' \quad (31)$$

$$\tan\theta = \frac{\gamma^{-1}\sin\theta'}{1 - \beta\cos\theta'} \quad (32)$$

$$t = \gamma t'(1 - \beta\cos\theta') = D_{-c,\theta'} t' \quad (33)$$

$$t = \gamma^{-1} t'(1 + \beta\cos\theta)^{-1} = D_{-c,\theta} t' . \quad (34)$$

The space-time diagram that works in the case is presented in Figure 5.

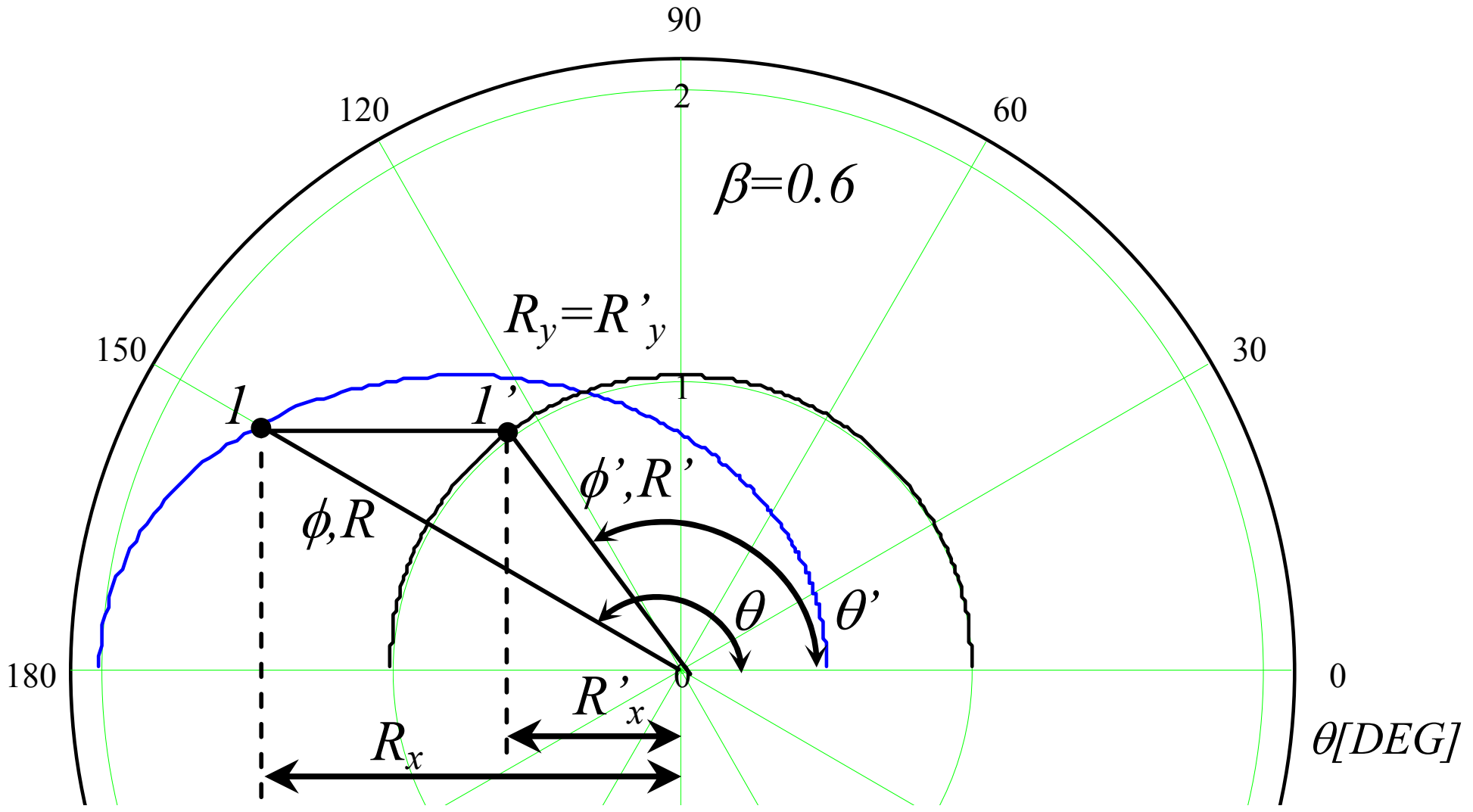

***Figure 5.*** *The relativistic diagram that displays in true values the magnitudes the space-time coordinates of events generated by light signals (photons) received at the origins of the two involved reference frames at t=t'=0. It also displays in true magnitudes the vector and the scalar components of four vectors that transform as the space-time coordinates of the events mentioned above do.*

It displays in true values, the magnitudes and the components of the four vectors. We measure on its axes the projections of the vector components. It also displays the circle of radius $R'(\Phi')$ and the curve described by (34). The invariance of the OY(O'Y') components $(R_y = R'_y)$ enables us to find out the correspondence between the components of the four vector as measured from K and K' respectively. It displays (in true values) the magnitudes and the components of the four vectors $(\mathbf{R}, \Phi)$ and $(\mathbf{R}', \Phi')$ associated with the approaching photon and with the events it generates that transform by definition as the space and the time coordinates do. In this case $(\mathbf{R}, \mathbf{R}')$ could represent the magnitudes of the position vector, photon momentum and wave vector, this diagram displaying theirs components as well. It also displays the magnitudes of the time coordinates, photon energy and frequency of the electromagnetic oscillations taking place in the wave that propagates towards the observer.

### 3. The space-time diagram at work

#### 3.1 Photographed and radar detected shape of a moving profile[8,9,10]

The radar detection is similar to the hunting with laser guns. If the radar detected profile is a circle at rest in K' then (as we have seen above) when we detect it from K its radar detected shape will be an ellipse. This ellipse has its left focal point at the origin O of the diagram when the detected profile is "receding", respectively its right focal point at O when the profile is "approaching" the observer O.

The photographic detection is similar to the simultaneous detection of light signals that have left a luminous profile at the different points and at different times. The time when the simultaneous detection takes place is the time when our observers $O_0$ and $O_0'$ open the photo cameras they handle or their eyes. If the photographed luminous object is a circle at rest in K' located at its origin O' then its photographed shape will be an ellipse. This ellipse has its right focal point at the origin O' of the diagram when the luminous circle is approaching the observer $O_0$, respectively the ellipse has its left focal point at the origin of the diagram when the circle is receding the observer $O_0$.

The results obtained so far tell us that the radar detected shape of a circle at rest in K' and having its centre at O' is an ellipse (13) having its left focal point at O' as well. They also tell us that the photographed shape of the same circle is an ellipse (31) having its right focal point at O' as well.

For illustrating more clearly the way in which our diagram works consider a profile at rest in K' described in polar coordinates by:

$$r' = F(a_1, a_2 ... a_n, \theta') \qquad (35)$$

where $a_1, a_2 ... a_n$ represent proper parameters and $\theta'$ represents the polar angle. If we detect the profile from K by sending light signals at $t' = 0$ in all directions then (18) tells us that the observers from K will describe its radar-detected shape by:

$$r = F(a_1, a_2 ... a_n, \theta')\gamma(1 + \beta\cos\theta'). \qquad (36)$$

If the profile is luminous and the observers from K receive light signals at t=0 that have left the profile at different times then its photographed shape is:

$$r = F(a_1, a_2 ... a_n, \theta)\gamma(1 - \beta\cos\theta') . \qquad (37)$$

Consider that the profile at rest in K' is a straight line parallel to the O'Y' axis located at a distance *a* apart from it. This line is described in K' by:

$$r = \frac{a}{\cos\theta'} . \qquad (38)$$

and its radar shape as detected from K is described by:

$$r = \gamma^{-1}a(\cos\theta - \beta)^{-1} \qquad (39)$$

while its photographic shape also as detected from K being described by:

$$r = \gamma^{-1}a(\cos\theta + \beta)^{-1} . \qquad (40)$$

In Figures 6 and 7 we present the way in which the relativistic diagram enables us to construct point-by-point the radar and respectively the photographed shape of the considered moving profile.

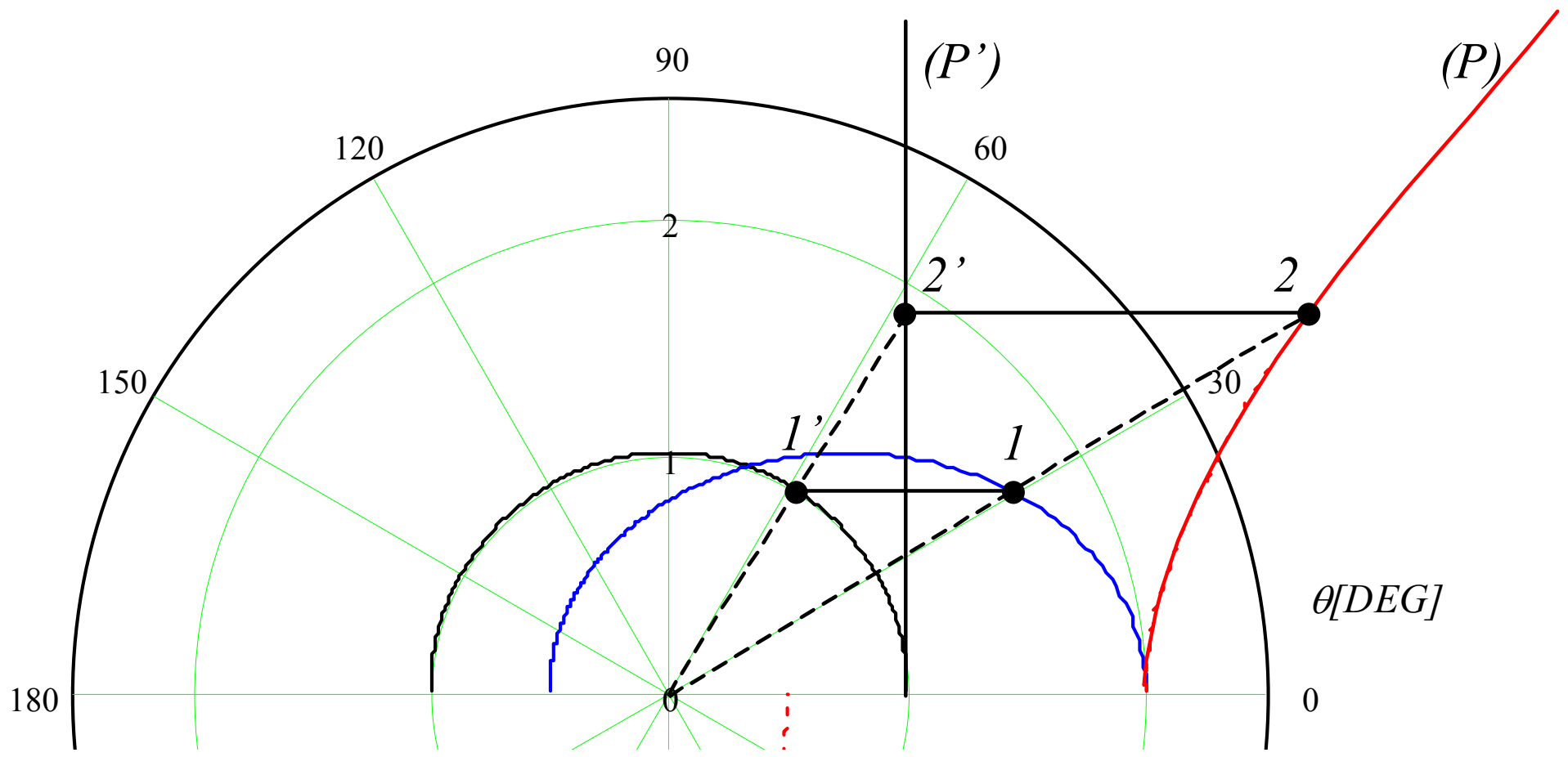

***Figure 6.*** *Illustrating the way in which the relativistic diagram enables us to construct the radar detected shape of a straight line at rest in K' and parallel to the O'Y' axis.*

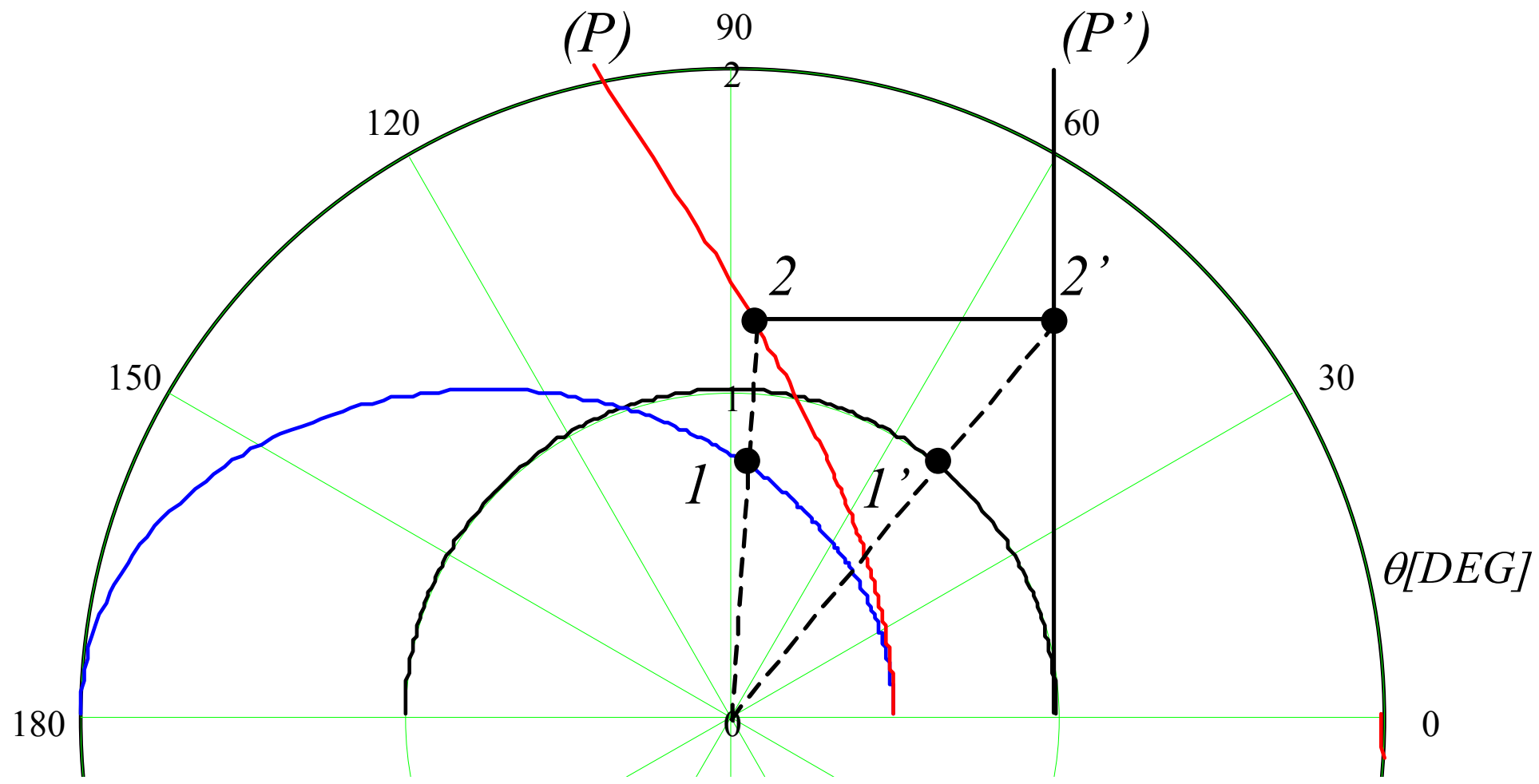

***Figure 7.*** *Illustrating the way in which our relativistic diagram enables observers from K to construct the photographed shape of a straight line at rest in K' parallel to the O'Y' axis.*

### 3.2 Photographed and radar detected length of a moving rod

Observers from K measure the length of a rod at rest in K'. The rod is parallel to the common axes OX(O'X'), at rest in K' where observers measure its proper length $L_0$. In Figures 8 and 9 we present the way in which the space-time diagrams enable us to measure its length from K using the radar and the photographic detection respectively. In both cases we have considered the situations when both ends of the rod are "receding" or "approaching" from the perspective of the stationary observer **O**(0,0). In both cases a length contraction or length dilation could take place or even we can detect (using and adequate location of the rod in its rest frame) the Lorentz contraction. An analytic approach is presented in [11].

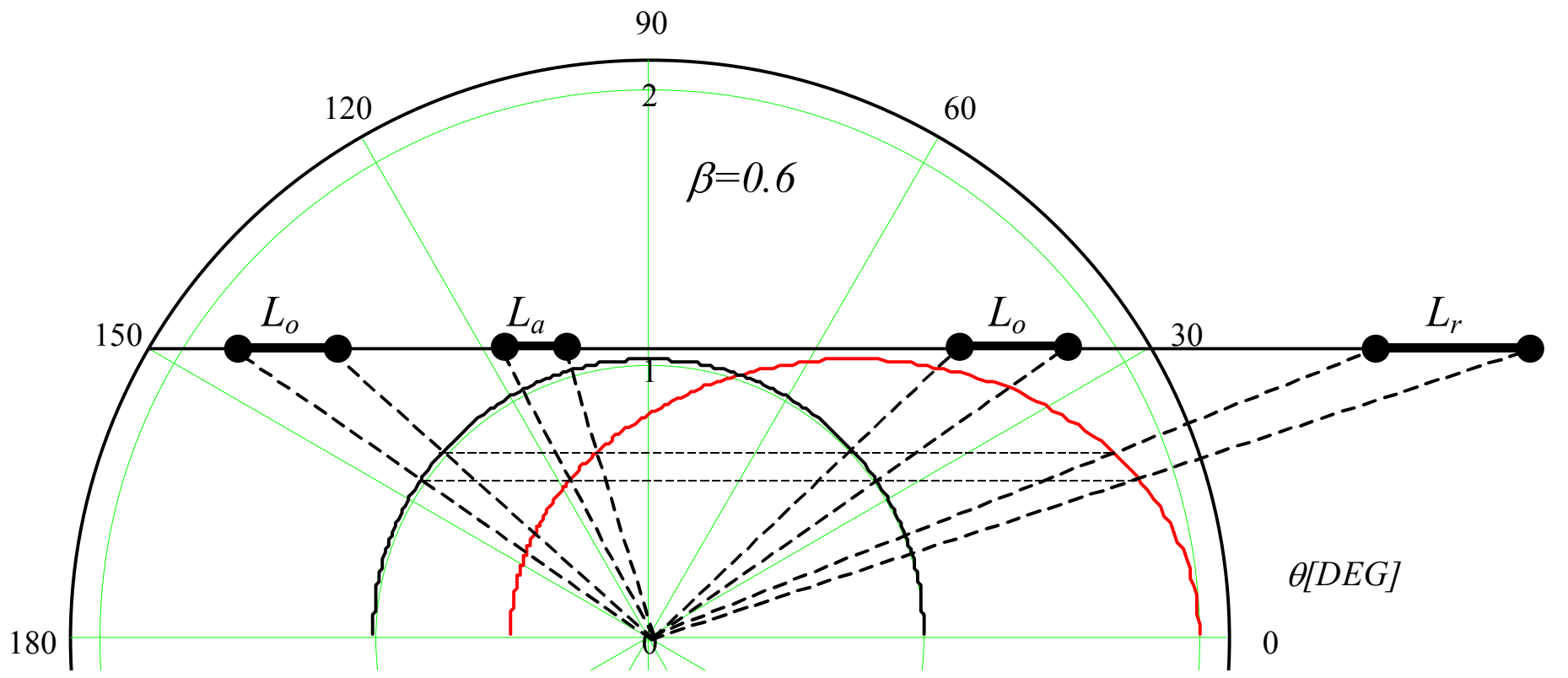

***Figure 8.*** *Illustrating the way in which observers from K measure the length of a rod of proper length $L_0$ at rest in K' and parallel to the common axes using the radar detection. Two cases are considered. In the first one both ends of the rod are approaching. In the second case both ends of the rod are receding the observer.*

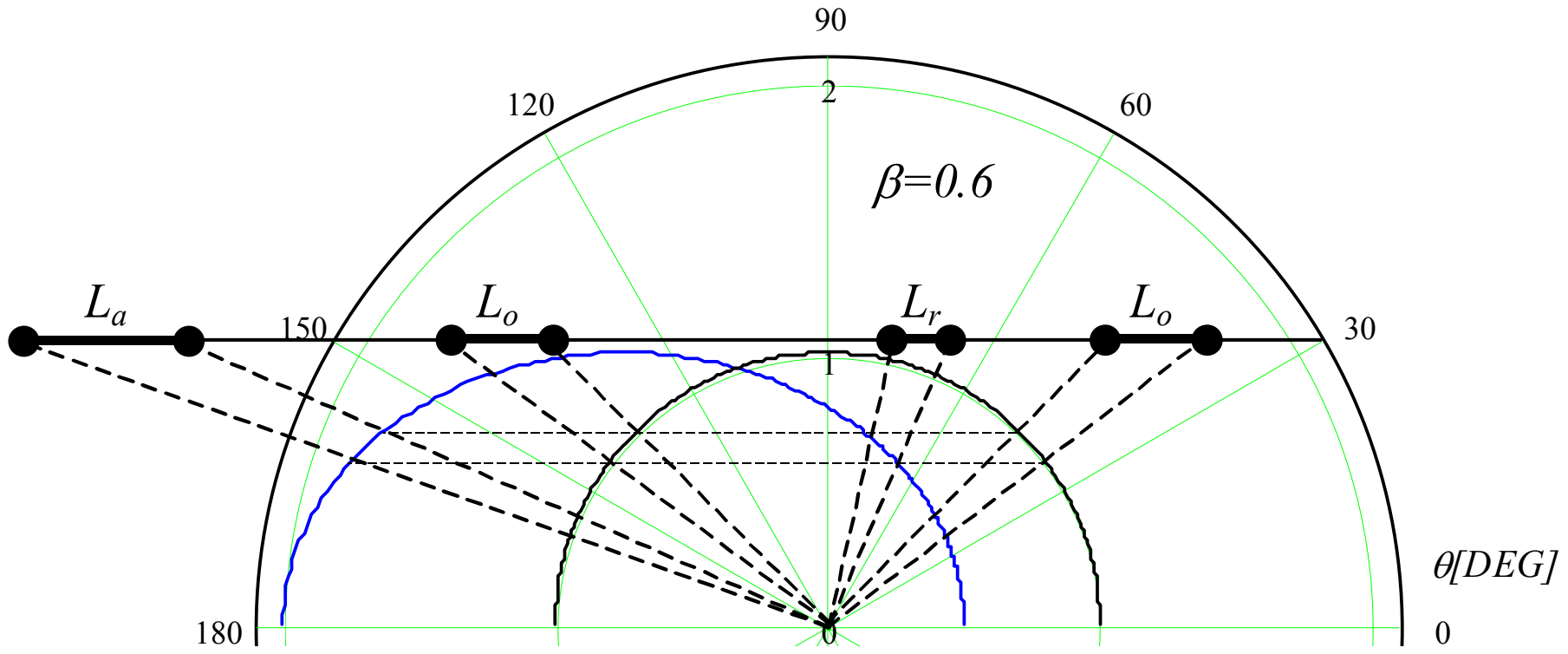

***Figure 9.*** *Illustrating the way in which observers from K measure the length of a rod of proper length $L_0$ at rest on K' and parallel to the common axes using photographing detection. Two cases are considered. In the first one bots ends of the rod are approaching. In the second case both ends of the rod are receding the observer.*

### 3.3 Einstein's mirror[12]

There are situations in which we should use both space-time diagrams (for incoming and outgoing light rays) at the same time. Consider a plane mirror at rest in K' and confined to the X'O'Z' plane. An incident ray makes an angle $\alpha$ with the normal to the mirror and it is reflected along a direction that makes the same angle with the normal as shown in Figure 10.

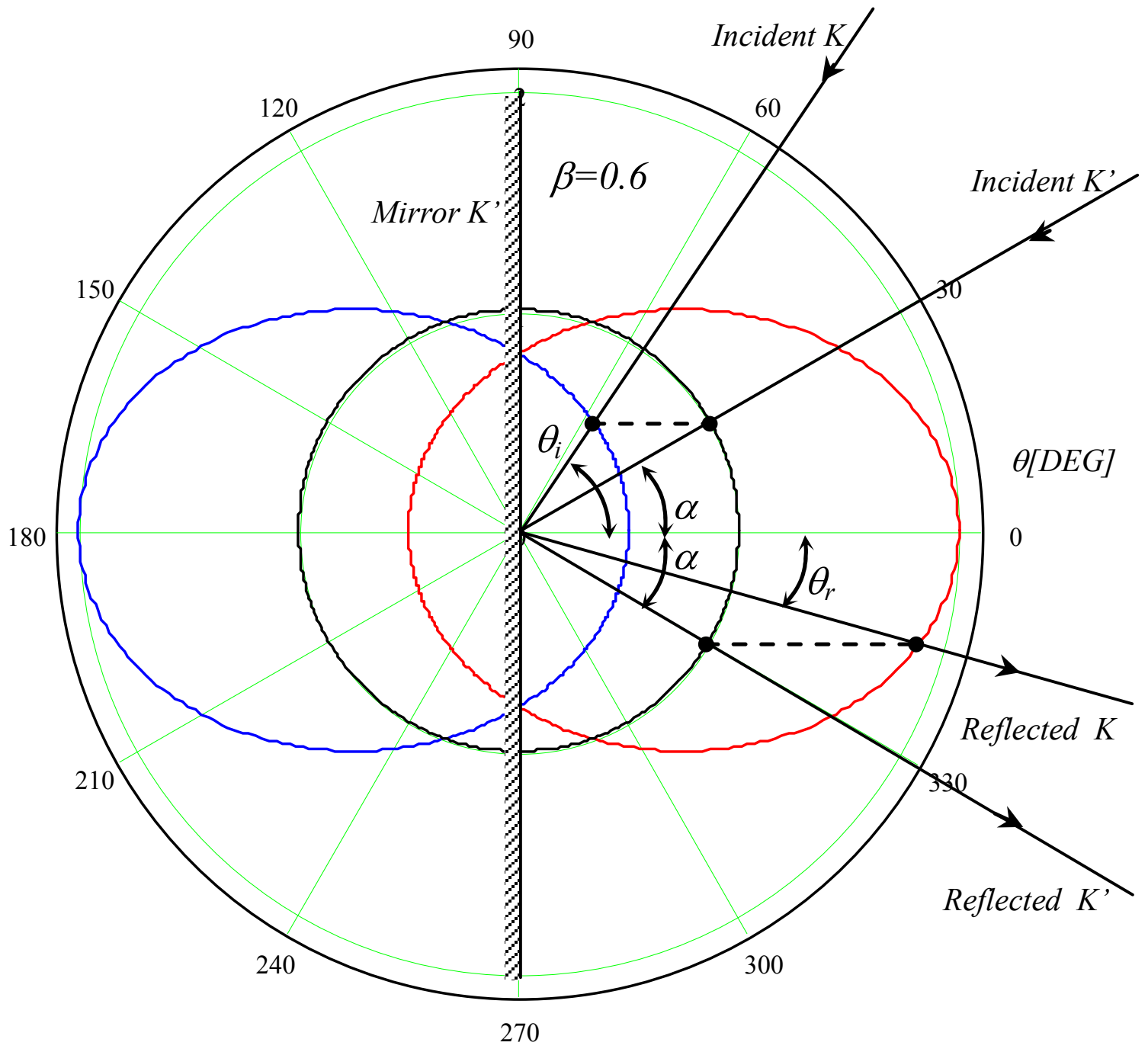

***Figure 10.*** *Illustrating the way in which observers from K detect the reflection of light on a vertical mirror at rest in K.*

On the same figure we present a circle of radius $\nu'$ (the frequency of the incident and the reflected rays as measured in K') at a given well-defined scale and the two space-time diagrams presented above. The reflection is considered on the right surface of the mirror.

Let 1' and 2' be two events detected from K' which take place on the circle, the first on the incident ray the second on the reflected one. According to the handling rules of the diagram, when detected from K the event 1' takes place on the ellipse corresponding to the incoming ray at point 1. Event 2' detected from K takes place at point 2 located on the ellipse that corresponds to the outgoing rays. The incidence point O that coincides with the origin of our space-time diagrams is the same in all inertial reference frames. Knowing the location of the points 1 and 2 we can draw the incident and the reflected rays as detected from K; the first ray makes an angle $i$ the second an angle $r$ with the normal to the mirror as detected from K'. All the angles are displayed in true values.

### 3.4 Privileged directions

Analyzing figures 2 and 3 we see that the basic curves of our space-time diagrams have two common points, determined by theirs intersection.

First consider the case of the four vectors associated with tardyons. Figure 2 accounts for the vector component of the four vector whereas Figure 3 accounts for its scalar component. In the case of the vector component the two points are $1(R,\theta)$ and $2(R,-\theta)$ when detected from K, respectively $1'(R'=R,\theta'=\pi-\theta)$ and $2'(R'=R',\theta'=-(\pi-\theta))$ when detected from K'. The particular direction $\theta$ is characterized by the fact that along it observers from K do not detect relativistic effects associated with the magnitude of ***R*** and ***R'***. Imposing the condition *R=R'* (4) leads to the following value for the angle $\theta$:

$$\cos\theta = \frac{1}{uVc^{-2}}\left(1-\sqrt{1-\frac{V^2}{c^2}}\right) \quad . \tag{41}$$

In the case of the scalar components ($\Phi,\Phi'$) (Figure 3), we see that the two points are $1(\Phi,\theta)$ and $2(\Phi,-\theta)$ when detected from K, respectively $1'(\Phi'=\Phi,\theta'=\pi-\theta)$ and $2'(\Phi'=\Phi,\theta'=-(\pi-\theta))$ when detected from K'. We obtain the value of the angle $\theta$ that characterizes the direction along which no relativistic effects are detected imposing in (6) the condition $\Phi=\Phi'$ that leads to the same relation as in (41).

We can do the same analysis for radar detection, the direction without relativistic effects being obtained in this case from (41) when *u=c* as

$$\cos\theta_c = \frac{1-\sqrt{1-\frac{V^2}{c^2}}}{\frac{V}{c}} \quad . \tag{42}$$

The space-time diagram accounting for the relativistic effects in the spherical wave illustrates the way in which observers from K detect blue-shift and red-shift ~~e~~ for the monochromatic radiation detected from K' where the frequency of the radiation is the same in all directions say $\nu'$, the space distribution being uniform. Detected from K this distribution is no longer uniform (see Figure 11).

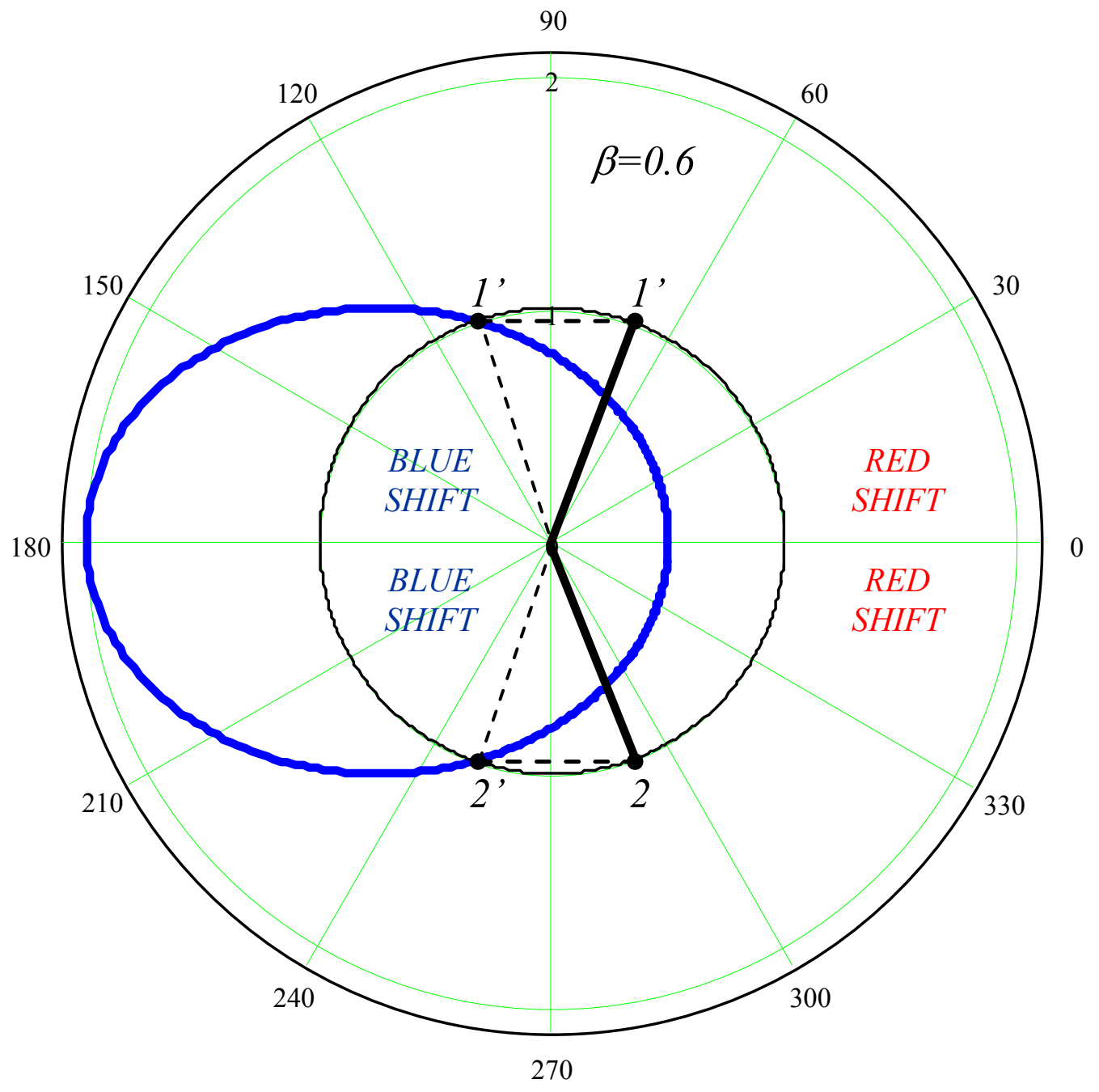

***Figure 11***. *Illustrating the directions along which no relativistic effects are detected for radar detection. Separating the regions within which a blue shift or a red shift takes place in the frequency of the electromagnetic radiation.*

For the directions $0<\theta<\theta_c$ we have $\nu>\nu'$ therefore a blue-shift takes place whereas for the directions $\theta_c<\theta<\pi$ we have $\nu<\nu'$ therefore a red-shift takes place. The probability to detect a blue shifted photon is:

$$w_b=\frac{\theta_c}{\pi} \tag{43}$$

while the probability to detect a red shifted photon is:

$$w_r=1-w_b\ . \tag{44}$$

In Figure 12 we present the red and blue shift in the case of the photographic detection.

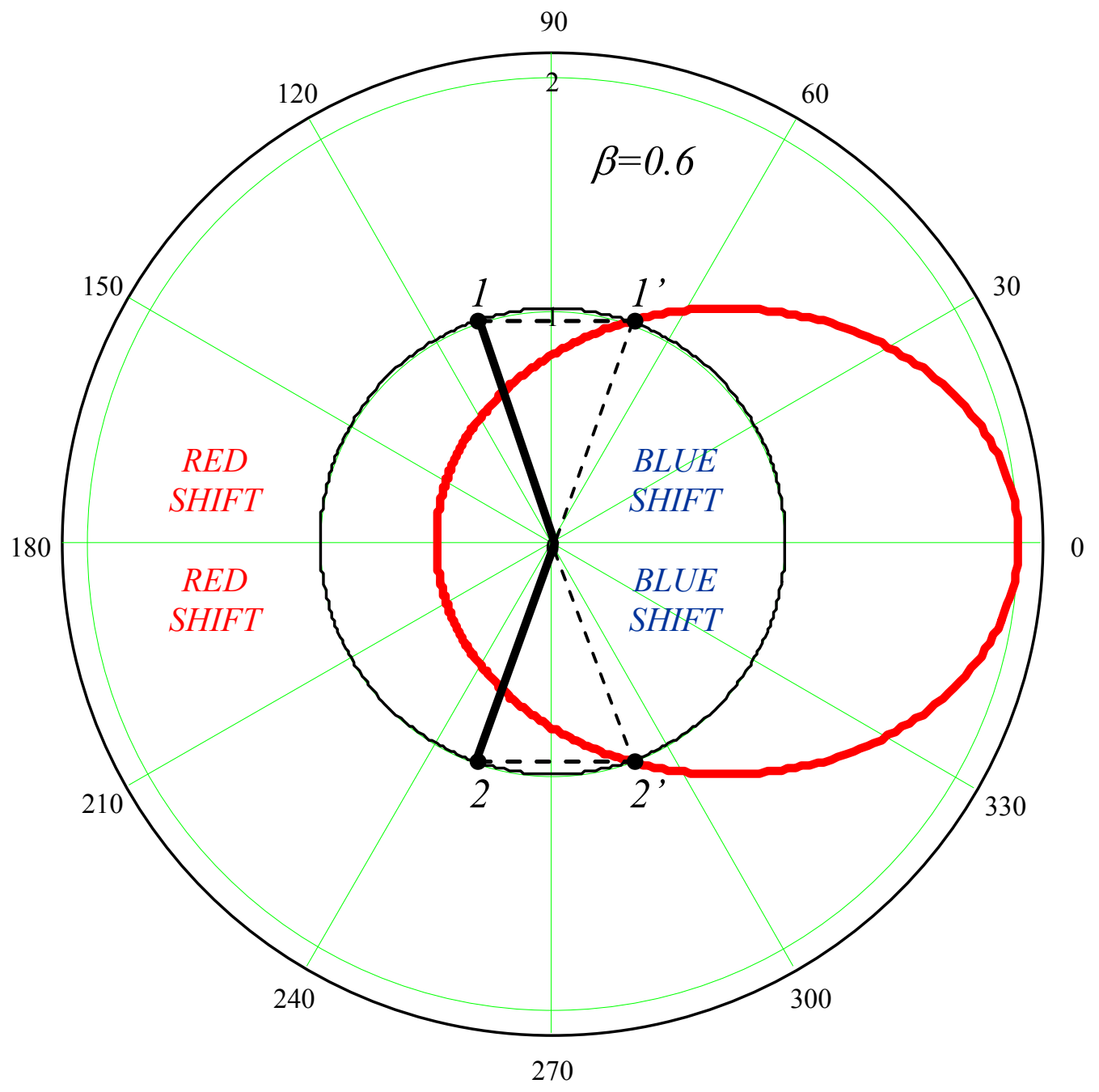

***Figure 12.*** *Illustrating the directions along which no relativistic effects are detected for the photographic detection. Separating the regions within which a blue shift or a red shift takes place in the frequency of the electromagnetic radiation.*

## 4. Conclusions

The relativistic diagram we propose displays in true magnitudes the vector and the scalar components of four vectors. These components transform as the space-time coordinates of events generated by moving tardyons or photons do.